\documentclass[]{achemso}
\setkeys{acs}{keywords = true}

\usepackage{lipsum} 
\usepackage[dvipsnames]{xcolor} 
\usepackage{graphicx} 
\usepackage{grffile}
\usepackage{amsmath, amssymb} 
\usepackage{bm} 
\usepackage{xr} 
\usepackage{makecell}
\usepackage{changepage}
\usepackage{pdflscape}
\usepackage{mathtools}
\usepackage[thinc]{esdiff}

\usepackage{pgffor}

\usepackage{wrapfig}

\usepackage{nameref}
\usepackage{varioref}
\usepackage{hyperref}
\usepackage{cleveref}

\usepackage{layout}
\usepackage{placeins}

\usepackage[normalem]{ulem} 

\usepackage{enumitem} 

\usepackage{siunitx} 

\usepackage[most]{tcolorbox} 

\title{Glassy interphases reinforce elastomeric nanocomposites by enhancing percolation-driven volume expansion under strain}
\author{Pierre Kawak}
\affiliation{Department of Chemical, Biological, and Materials Engineering, University of South Florida, Tampa, Florida 33612}
\author{Harshad Bhapkar}
\affiliation{Department of Chemical, Biological, and Materials Engineering, University of South Florida, Tampa, Florida 33612 }
\author{David S.~Simmons}
\affiliation{Department of Chemical, Biological, and Materials Engineering, University of South Florida, Tampa, Florida \ 33612 }
\date{\today}
\email{dssimmons@usf.edu}

\keywords{elastomer nanocomposites; filled elastomers; nanoparticle reinforcement; granular particle networks; glassy interphase; bound rubber; Poisson's ratio mismatch; bulk‑modulus contribution; low‑strain reinforcement; jamming; percolation; molecular simulations}
\newcommand*{\abstracttext}{
  For nearly a century, introduction of nanoparticles to elastomers has yielded extraordinarily tough nanocomposites that are critical to technologies from actuators to tires. The mechanisms by which this reinforcement occurs have nevertheless remained a central open question in material science. One widely debated hypothesis posits that strong interactions between polymer and particles induce ``glassy bridges'' that cement particles into a cohesive percolating network that resists elongation. Here, molecular dynamics simulations show that glassy particle shells do not primarily provide elongational cohesion. Instead, they amplify an underlying mechanism wherein competition between filler and elastomer networks causes the elastomer's volume to increase on deformation. This induces contributions from the elastomer's bulk modulus, which is of order 1000 times larger than its Young's modulus. These findings establish a unified understanding of low-strain reinforcement in filled elastomers as emanating from volumetric competition between coexisting particulate and elastomeric networks. This reframes and unifies our understanding of low-strain reinforcement, provides a clear-cut diagnostic for the presence of glassy bridging, and offers a new design principle for tough elastomeric nanocomposites.
}

\makeatletter
\renewcommand*{\acs@maketitle@extras}{%
  \acs@maketitle@extras@hook
  \newpage
}
\makeatother
\let\oldmaketitle\maketitle
\let\maketitle\relax

\begin{document}

\begin{@twocolumnfalse}
  \oldmaketitle

  \begin{abstract}
    \abstracttext
  \end{abstract}

\begin{tcolorbox}[
  colback=gray!5,              
  colframe=black,              
  colbacktitle=black,          
  coltitle=white,              
  title={\large\bfseries Significance Statement}, 
  fonttitle=\sffamily,         
  boxrule=0.8pt,
  arc=3mm,
  left=5mm, right=5mm,
  top=3mm, bottom=3mm,
  enhanced,
  titlerule=0pt,
  toptitle=2mm, bottomtitle=2mm 
]
By adding nanoparticles to elastomers, engineers turn materials with the give of rubber bands into composites strong enough to support a Boeing 747 on just a few tire patches. The physics driving this transformation has remained elusive for nearly a century. Using molecular simulations, we show that nanoparticles primarily reinforce elastomers by forcing them to expand in volume under strain, drawing on the polymer’s incompressibility to enhance its resistance to deformation. When nanoparticles are sticky, they cause surrounding elastomer layers to become viscous or even glassy, further amplifying this effect.  These findings unify competing hypotheses for the mechanism into a cohesive model of how nanoparticles strengthen elastomers.
\end{tcolorbox}

\end{@twocolumnfalse}

\section{Introduction}


Engineers have long added nanofillers, such as carbon black or silica, to polymeric elastomers in order to make them stiffer and tougher~\cite{Robertson2021} -- enhancements that enable rubber to withstand millions of cycles of deformation without failing (e.g. in vehicle tires, vibration dampers, and actuators). 
The surface of the nanoparticles in these systems are often `sticky' -- that is, they can attract and immobilize nearby polymer segments.
This stickiness is believed to play a central role in the mechanical performance of these~\cite{qu_nanoscale_2011, fleck_polymer-filler_2014} and other nanocomposites\cite{cheng_unraveling_2016, Huang2022, bokobza_elastomer_2023}.
However, the mechanism of this effect, and even the extent to which it genuinely drives nanoparticulate reinforcement of elastomers remain unresolved. The central challenge is to disentangle multiple physical mechanisms that may arise in the presence of attractive nanoparticles (strong polymer-particle interactions)~\cite{cheng_focus_2017}. This represents a nearly 100-year problem in soft materials science: despite elastomeric nanocomposites' role as the oldest and most economically-important nanocomposite, we do not understand their fundamental behavior.

The large reinforcements realized in experimental nanocomposites (e.g. 10$\times$ or greater modulus enhancements) are now known to result from mechanical percolation of nanoparticles.~\cite{Pryamitsyn2006, Steck2023, Boschan2016, Huang2022, Simmons2016, Wang2005, Richter2011, Roland2015} However, the nature of this percolation and its relationship to polymer-particle interactions remains an open question. Do particles percolate through direct particle-particle contacts?~\cite{Krishnamoorti1997, Krishnamoorti2001, Galgali2001,Witten1993} In this case, what role do strong polymer-particle interactions play? Are particles instead interconnected via "bridges" of immobilized, glassy polymeric segments induced by strong polymer-particle interactions?~\cite{Berriot2002, Berriot2003, Merabia2008, Montes2003, Mujtaba2014, chen_mechanical_2015}  If so, how can we account for reinforcement in systems apparently lacking such effects?~\cite{kumar_50th_2017,Robertson2008,Robertson2008a,Smith2017,Smith2019,Kawak2024}

Here we employ molecular dynamics (MD) simulations of elastomeric nanocomposites incorporating strong polymer-particle attractions (strength controlled by parameter $\epsilon_{PF}$) in order to answer these questions.
To do this, we explore the synergistic effects of $\epsilon_{PF}$ with filler loading $\phi_{F}$ and structure $N_{p}$ on various mechanisms of reinforcement by measuring relaxation times, bulk moduli, Young's moduli, and Poisson's ratios for neat and filled elastomers.
In doing so, we seek to reconcile several sets of facts.
First, polymer dynamics can indeed be significantly slowed in the vicinity of attractive, rigid surfaces, reflecting a local increase in the glass transition temperature $T_g$.~\cite{simmons_emerging_2016, Berriot2003, Starr2016, Ondreas2019, Schweizer2019, Simmons2016, bindu_viscoelastic_2013, alcoutlabi_effects_2005,forrest_glass_2001,roth_glass_2005,mckenna_ten_2010,richert_dynamics_2011,ediger_dynamics_2014,broth_polymers_2021, ruan_glass_2015,fryer_dependence_2001,lee_correspondence_2017,Lang2014,gao_existence_2014,DiazVela2020,Fakhraai2008,Daley2012}. This makes the glassy bridge scenario highly plausible. On the other hand, this proposal remains highly debated, with a distinct set of studies questioning its strength of evidence and relevance at typical use temperature far above $T_g$.~\cite{bogoslovov_effect_2008,Warasitthinon2018, Robertson2008a,Robertson2008,Robertson2011,kumar_50th_2017,Kawak2025}
Indeed, early work by Witten, Rubinstein and Colby established a mechanism by which reinforcement could instead occur via direct particle-particle contacts, via resistance to lateral compression during deformation.~\cite{Witten1993} Our recent simulation work has shown that a compression-dominated scenario reminiscent of this picture indeed governs reinforcement in the absence of glassy bridge effects.\cite{Smith2017,Smith2019,Kawak2024,Kawak2025}

With this in mind, we seek to disentangle four possible mechanisms in which strong polymer-particle attractions could hypothetically augment mechanical response.
\begin{enumerate}[wide, label=M\arabic*)]
\item \emph{Strain localization}: strong attractions to particles could immobilize surrounding polymer, leading to strain localization and stress amplification in the remaining mobile elastomer domains. This ``bound-rubber" mechanism was popular in early literature~\cite{bogoslovov_effect_2008}.
\item \emph{Glassy bridging}: regions of polymer between particles may vitrify to form glassy bridges, yielding a collective elongational cohesion of the nanoparticle network.~\cite{Huang2022, Dannenberg1986, Berriot2002, Berriot2003, Montes2003, Merabia2008, Papon2012, Mujtaba2014, chen_mechanical_2015, Starr2016, Sotta2017, Tauban2017} 
\item \emph{Transient crosslinking}: arrested or dynamically slowed polymer regions around particles, or adhesions to the particles themselves, may act as long-lived physical crosslinks in the matrix.
This could increase the effective crosslink density of the rubber, thus increasing the entropic elastic modulus of the polymer domains.~\cite{Sternstein2002, Steck2023, Dhara2022, Dhara2024, berriot_fillerelastomer_2002}
\item \emph{Enhanced Poisson's ratio mismatch effects}. Finally, reduced-mobility polymer (glassy shells) around nanoparticles could enhance mechanical response by augmenting direct-contact reinforcement effects. Our recent work has shown that reinforcement in systems lacking glassy bridge effects emerges from a polymer-particle competition over the \emph{volume} of the system under deformation.~\cite{Smith2017,Smith2019,Kawak2024,Kawak2025} Whereas elastomers tend to conserve volume under deformation (corresponding to a Poisson's ratio of 0.5), jammed particle networks tend to increase in volume under deformation (corresponding to a Poisson's ratio in the vicinity of 0.3). Because these networks share a boundary condition, the composite exhibits intermediate behavior, leading to a growth in volume on deformation of the composite. This is resisted by the ($\mathcal{O}\left(\SI{}{\giga\pascal}\right)$) bulk modulus of the elastomer matrix, dramatically enhancing resistance to deformation.~\cite{Smith2017,Smith2019, Kawak2024, Kawak2025}. We expect that this effect will intensify in the presence of glassy shells that augment particle jamming behavior.
\end{enumerate}

Our findings in this work show that while all of these mechanisms play a role in reinforcing the material, the primary impact of strong polymer-particle attraction is to enhance the Poisson's ratio mismatch (M4) in the filled elastomer.
In other words, the dominant reinforcement arises not from glassy shell-mediated elongational cohesion, but from a mismatch in Poisson's ratio that draws a large bulk modulus contribution into the material stiffness.
We also find that glassy shells around nanoparticles introduce a distinct ultra-low-strain ($<2\%$) response regime (M2).
Our findings also point to a diagnostic signature for glassy bridges based on the presence of an ultra-low-strain glassy response regime.
Taken together with our prior work, this study offers a unified and comprehensive framework for understanding nanoparticle-induced reinforcement in elastomers from low-strain deformation all the way to fracture.


\section{Methods} \label{sec:methods}


We perform molecular dynamics simulations of elastomeric nanocomposites under uniaxial extension, using the Large-scale Atomic/Molecular Massively Parallel Simulator (LAMMPS) package.~\cite{Intveld2008, Brown2011, Thompson2022}
Simulations employ a timestep of $10^{-3}\tau_{\text{LJ}}$, with $\tau_{\text{LJ}}$ representing the Lennard-Jones (LJ) time unit, approximately equivalent to one picosecond. Simulations under deformation employ a constant engineering strain rate of $5 \times 10^{-5}/\tau_{\text{LJ}}$.
This strain rate is selected based on prior studies that confirm it falls within the rubbery-plateau regime for stress response in our baseline neat crosslinked polymer model.~\cite{Smith2017,Smith2019}

We compute stresses locally within the system, both for different molecular species and in a spatially-resolved manner, to elucidate the reinforcing mechanisms inherent to these nanocomposites.
Two types of model systems are studied: (1) an unfilled polymer network composed of 5000 cross-linked polymer strands, each strand consisting of 20 Kremer-Grest (KG) beads, with a cross-linking density of 95\%; and (2) a set of filled polymer systems in which highly \emph{dispersed} filler clusters are introduced to form a composite.
System set (2) is characterized by the filler volume fraction ($\phi_{F}$), the extent of structuring of each nanoparticle cluster as defined by a count of primary icosahedral particles from which it is constructed ($N_p$), and the strength of the polymer-filler interaction ($\epsilon_{PF}$).
Three distinct filler volume fractions are considered: 75, 100, and 125 parts per hundred rubber (phr), equivalent to volume fractions $\phi_{F}$ of 0.261, 0.320, and 0.371, respectively.
Additionally, filler cluster sizes $N_{p}$ ranging from 3 to 11 particles are explored.
Finally, we explore how the polymer-filler attraction strength impacts the deformation with a range of $\epsilon_{PF}$ values from 1 to 5.

\subsection{Coarse-grained polymer model}


The polymer network consists of 5000 polymer chains, each comprised of 20 beads interacting via an variant of the Kremer-Grest (KG) bead-spring model, and modified to include attractions, allow bond breakage, and include end-cross-linking.~\cite{Duering1994,Svaneborg2005}
Cross-linking is introduced by adding 2500 cross-linker beads that form bonds with polymer end beads within a distance of $1.3\sigma_{LJ}$, stopping after 95\% cross-linking is achieved.
Nonbonded interactions between beads are modeled using a truncated Lennard-Jones (LJ) potential:

\begin{equation}
        E_{\text{LJ}}(r) = \begin{cases}
                4\epsilon\left[\left(\frac{\sigma_{LJ}}{r}\right)^{12} - \left(\frac{\sigma_{LJ}}{r}\right)^6\right] & r < r_c \\
                0 & r \geq r_c
        \end{cases}
\end{equation}
where $\epsilon$ and $\sigma_{LJ}$ are the characteristic energy and length scales, respectively, and the cutoff radius $r_c$ is set to $2.5\sigma_{LJ}$.
Bonded beads interact via a breakable quartic potential,
\begin{equation}
        E_{\text{bond, quartic}}(r) = k_q (r - r_0)^2 (r - r_0 - B_1) (r - r_0 - B_2) + E_0 + E_{\text{LJ}}(r) + \epsilon \ ,
\end{equation}
although no bonds break at the low strains employed in the present simulations. The parameters for the polymer bonding potential are chosen to match the shape of the standard KG FENE potential at energies (bond length fluctuations) appreciably sampled in quiescent simulations.
Specifically, the spring constant is $k_{q}=2351$, maximum extensibility before bond failure is $r_0=1.5$, distance parameters are $B_1=-0.7425$ and $B_{2}=0$, and energy the parameter is $E_0=92.74$.
This parameterization is also summarized in Table~\ref{tab:quartic_params}.


\subsection{Modeling nanofiller clusters}


Filler clusters are constructed from primary particles, each comprising a central bead surrounded by three concentric shells of beads (inner, middle, outer), totaling 147 beads per primary particle.
Clusters are formed by sintering together between 3 to 11 primary particles ($N_p$), randomly arranged.
Filler particles interact via the same nonbonded potential but with a truncated $r_c=2^{1/6}$, rendering their self interaction purely repulsive.
Bonds within and between shells are described using a quartic potential, with parameters detailed in Table~\ref{tab:quartic_params}.

\begin{table}[ht]
        \centering
        \caption{Quartic bond parameters for polymer and filler bonds. For filler bonds, interactions are specified according to shell types (center, inner, middle, outer).}
        \label{tab:quartic_params}
        \begin{tabular}{lcccccc}
                \hline
                Bond Type             & $r_{eq}$ & $k_q$ & $B_1$ & $B_2$ & $r_0$ & $E_0$ \\
                \hline
                Polymer               & 0.9609 & 2351.00 & -0.7425 & 0.0000 & 1.5000 & 92.74 \\
                Center-Inner          & 0.9511 & -4513.41 & -1.270 & -0.9920 & 1.4902 & 451.5 \\
                Inner-Middle (near)   & 0.9511 & -4513.41 & -1.270 & -0.9920 & 1.4902 & 451.5 \\
                Middle-Outer (near)   & 0.9511 & -4513.41 & -1.270 & -0.9920 & 1.4902 & 451.5 \\
                Sinters (near)        & 1.0000 & -5765.89 & -1.118 & -1.0560 & 1.5391 & 520.6 \\
                Inner-Inner           & 1.0000 & -5765.89 & -1.118 & -1.0560 & 1.5391 & 520.6 \\
                Middle-Middle         & 1.0000 & -5765.89 & -1.118 & -1.0560 & 1.5391 & 520.6 \\
                Outer-Outer           & 1.0000 & -5765.89 & -1.118 & -1.0560 & 1.5391 & 520.6 \\
                Inner-Middle (cross)  & 1.3800 & -7166.46 & -1.030 & -1.1359 & 1.9191 & 629.8 \\
                Middle-Outer (cross)  & 1.3800 & -7166.46 & -1.030 & -1.1359 & 1.9191 & 629.8 \\
                Sinters (cross)       & 1.4142 & -6610.81 & -1.015 & -1.1598 & 1.9533 & 587.1 \\
                \hline
        \end{tabular}
\end{table}


\subsection{Controlling polymer-filler interaction}


The polymer-filler interaction strength $\epsilon_{PF}$ is varied systematically as a multiple of the polymer-polymer interaction strength $\epsilon_{PP}$, such that $\epsilon_{PF}=\gamma\epsilon_{FF}=\gamma\epsilon_{PP}$, where $\gamma$ ranges from 1 to 5.
Adjusting $\epsilon_{PF}$ (or otherwise changing the strength of polymer-particle attractions) influences the local mobility of polymer segments near filler surfaces, as demonstrated by a range of prior simulations~\cite{hanakata_local_2012, Lang2014, starr_modifying_2011,Starr2016}.


\subsection{Filler dispersion and in-situ crosslinking} \label{sec:methods:anneal}


Filler clusters are dispersed into the polymer melt prior to crosslinking using a multi-step annealing protocol combining soft nonbonded potentials with gradual densification through successive NPT and NVT ensemble MD simulations, optimizing filler dispersion to avoid overlaps or agglomeration.
The process begins with populating a system of polymer chains and crosslinker beads with the desired fraction of filler using Packmol.~\cite{Martinez2009}
This detailed annealing and crosslinking procedure is fully outlined in Table 3 of reference~\citenum{Kawak2024}.
This protocol ends with a long equilibration period to minimize residual stresses.


\subsection{LAMMPS uniaxial extension simulations and stress determinations} \label{sec:methods:stretch}


Uniaxial deformation simulations are conducted at a strain rate of $5\times10^{-5}/\tau_{\text{LJ}}$ in the longitudinal direction, while maintaining zero normal pressure in transverse directions to allow deformation at the material's natural Poisson's ratio.
Pressure is maintained via an anisotropic Nos\'e-Hoover barostat with a damping parameter of $2\tau_{\text{LJ}}$.
Similarly, a unit temperature in LJ units is maintained via a Nos\'e-Hoover thermostat with a damping parameter of $5\tau_{\text{LJ}}$.
Furthermore, we mitigate the effects of nonphysical drift associated with momentum buildup due to round-off errors by zeroing out the linear momentum of the center of mass of all atoms every $1\times 10^4$ MD steps.

To compute stresses, we begin by computing a per-particle (non-normalized) stress contribution $s_{i}^{\alpha\beta}$ as the sum of kinetic energy and virial contributions for bead $i$, given by:
\begin{equation}
        s^{\alpha\beta}_{i} = - m v_i^\alpha v_i^\beta - W_i^{\alpha\beta},
\end{equation}
where $W_i^{\alpha\beta}$ is the Virial for bead $i$, $m_i$ is the mass of particle $i$, and $v_i^\alpha$ and $v_i^\beta$ are the $\alpha$ and $\beta$ components of particle $i$'s velocity.
Components of the system's total engineering stress are then computed as
\begin{equation} \label{eq:stress}
        \sigma^{\alpha\beta} = \frac{1}{\left(1+\varepsilon\right)V_{0}} \sum_{i=1}^{N} s_i^{\alpha\beta}
\end{equation}
where $\alpha$ and $\beta$ are the $x$, $y$, or $z$ dimensions, $\varepsilon$ is the extensional strain, and $V_{0}$ is the initial total volume. We define a per-species engineering stress of species $k$ as
\begin{equation} \label{eq:stress-species}
        \sigma^{\alpha\beta}_{k} = \frac{1}{\left(1+\varepsilon\right)V_{0}} \sum_{i=1}^{N_k} s_{k,i}^{\alpha\beta}
\end{equation}
where $N_k$ is the number of beads of species $k$, such that the sum is over only beads of species $k$.

We note that the above definition of a partial per-species stress parallels the standard definition of a partial pressure, in that per-species stresses are additive to the whole system stress:
\begin{equation} \label{eq:componentstresssum}
        \sigma^{\alpha\beta} = \sum_{k=1}^{K} \sigma^{\alpha\beta}_k
\end{equation}
where $K$ is the number of species in the system. However, it should be noted that, because of our use of the engineering (rather than true) stress, the stresses computed here do not correspond to \emph{thermodynamic} pressure components, except in the very low strain limit.

It is in many cases useful to define a per-species engineering stress that is normalized \emph{by its own partial volume} $V_{k,0}$ rather than by the total volume of the system. This is relevant when comparing per-component properties to their neat equivalents. We denote this quantity the partial volumetric component stress, $\bar{\sigma}^{\alpha\beta}_{k}$, defined as follows:
\begin{equation} \label{eq:stress-bar-species}
        \bar{\sigma}^{\alpha\beta}_{k} = \frac{1}{\left(1+\varepsilon\right)V_{k,0}} \sum_{i=1}^{N_k} s_{k,i}^{\alpha\beta} \ .
\end{equation}
This quantity relates to the component stresses and total stress via the following equations:
\begin{equation} \label{eq:stress-interrelation}
        \sigma^{\alpha\beta}_{k} = \phi_{k,0} \bar{\sigma}^{\alpha\beta}_{k}
\end{equation}
and
\begin{equation} \label{eq:volumetric-stress-sum}
        \sigma^{\alpha\beta} = \sum_{k=1}^{K} \phi_{k,0} \bar{\sigma}^{\alpha\beta}_{k}
\end{equation}
where $\phi_{k,0}$ is the initial volume fraction of component $k$ at zero strain.
In this work, we only utilize the initial volume in computing $\phi_{k}$ and drop the $0$ subscript in referring to initial volume fractions hereafter for brevity. To reduce statistical noise associated with stress measurements, all stresses are recorded at each MD time step and then time averaged.

We employ a notation for engineering modulus and partial engineering moduli that is consistent with the notation above.
The total system's engineering elongational tangent modulus, at an engineering strain $\varepsilon$ in the elongational direction, is given by
\begin{equation} \label{eq:total-modulus-definition}
        E(\varepsilon)={\left. \frac{d \sigma^{xx}}{d \varepsilon}\right|}_{\varepsilon}
\end{equation}
where $x$ is the direction of elongation.
Component moduli are defined as
\begin{equation} \label{eq:component-modulus-definition}
        E_k(\varepsilon)={\left. \frac{d \sigma_k^{xx}}{d \varepsilon}\right|}_{\varepsilon}
\end{equation}
such that
\begin{equation} \label{eq:modulus-sum}
        E=\sum_{k=1}^{K} E_{k}
\end{equation}
Partial volumetric component moduli are defined as
\begin{equation} \label{eq:component-modulus-bar-definition}
        \bar{E}_k(\varepsilon)={\left. \frac{d \bar{\sigma}_k^{xx}}{d \varepsilon}\right|}_{\varepsilon}
\end{equation}
such that
\begin{equation} \label{eq:modulus-bar-sum}
        E=\sum_{k=1}^{K} \phi_{k,0}\bar{E}_{k}
\end{equation}

We compute the Poisson's ratio $\nu$ by measuring the transverse strains relative to the applied longitudinal strain at a strain of 5.5\%, using finite differences to represent the derivative in Equation~\ref{eq:poisson}.
\begin{equation} \label{eq:poisson}
    \nu \equiv - \frac{d \left( \log L_{y} \right)}{d \left( \log L_{x} \right)}
\end{equation}


\subsection{Quantifying polymer segment dynamics via self-intermediate scattering functions} \label{sec:methods:isfs}


Polymer segmental dynamics near and far from fillers are quantified by fitting time-resolved self-intermediate scattering functions computed at a wavenumber of 7.07 (approximately the location of first peak in structure factor) to a two-exponential Kohlrausch-Williams-Watts (KWW) form.
The relaxation time in this work is defined as the time at which the value of the KWW fit decaying to a value of 0.2. This protocol is consistent with many prior simulation works probing surface gradients in dynamics.~\cite{hanakata_local_2012,lee_correspondence_2017,Lang2014, lang_interfacial_2013, ghanekarade_signature_2023, ghanekarade_nature_2021}
We characterize polymer segments by their nearest proximity to a filler surface bead as part of the first, second, third, or fourth shell around a filler particle with each shell having a thickness of $\sigma_{LJ}$.
In the text, we refer to the first shell as the interfacial polymer or the interface and the fourth shell as the bulk polymer (which is good to a reasonable approximation at the high reduced temperature employed in these simulations; at lower temperatures the effective range of altered near-interfacial dynamics increase~\cite{lang_interfacial_2013, Schweizer2019}).
Furthermore, polymer segments positioned between distinct filler clusters are identified and analyzed separately, providing insight into polymer-mediated interactions between filler particles, such as glassy or viscous bridges.
This region, the interfiller polymer, is identified via an algorithm described in our former work based on geometric overlap regions around icosahedral centers,~\cite{Kawak2024} with a characteristic distance of $17/8\sigma_{LJ}$.


\subsection{Computing bulk modulus of the neat elastomer} \label{sec:methods:bulk}


We compute the bulk modulus of the neat elastomer using the average and variance of the volume from an NPT ensemble MD simulation at unit temperature and zero pressure.
Specifically, we employ the fluctuation-dissipation relationship for the bulk modulus, $K$:~\cite{Hansen2006}
\begin{equation} \label{eq:bulk}
        K = T \frac{ \langle V \rangle }{ \langle \delta V ^2 \rangle }
\end{equation}
where $T$ is the temperature (in LJ units), $\langle V \rangle$ is the mean of the volume, and $\langle \delta V ^2 \rangle$ is the variance of the volume.
To use as an input to the Poisson's ratio mismatch theory relationship, we use this relationship on the NPT data for a neat elastomer to obtain $K_{e}$, where the subscript $e$ is used to denote the properties of the neat elastomer.
We use five replicates with $2 \times 10^{7}$ MD steps and a time step of $1 \times 10^{-3}$ in LJ units to obtain $K_{e} = 16.96 \pm 0.05$.



\section{Results}


\subsection{Glassy bridge and shell phenomenology in the quiescent limit} \label{sec:results:relax_time}

\begin{figure}[!htp]
  \centering
  \includegraphics[]{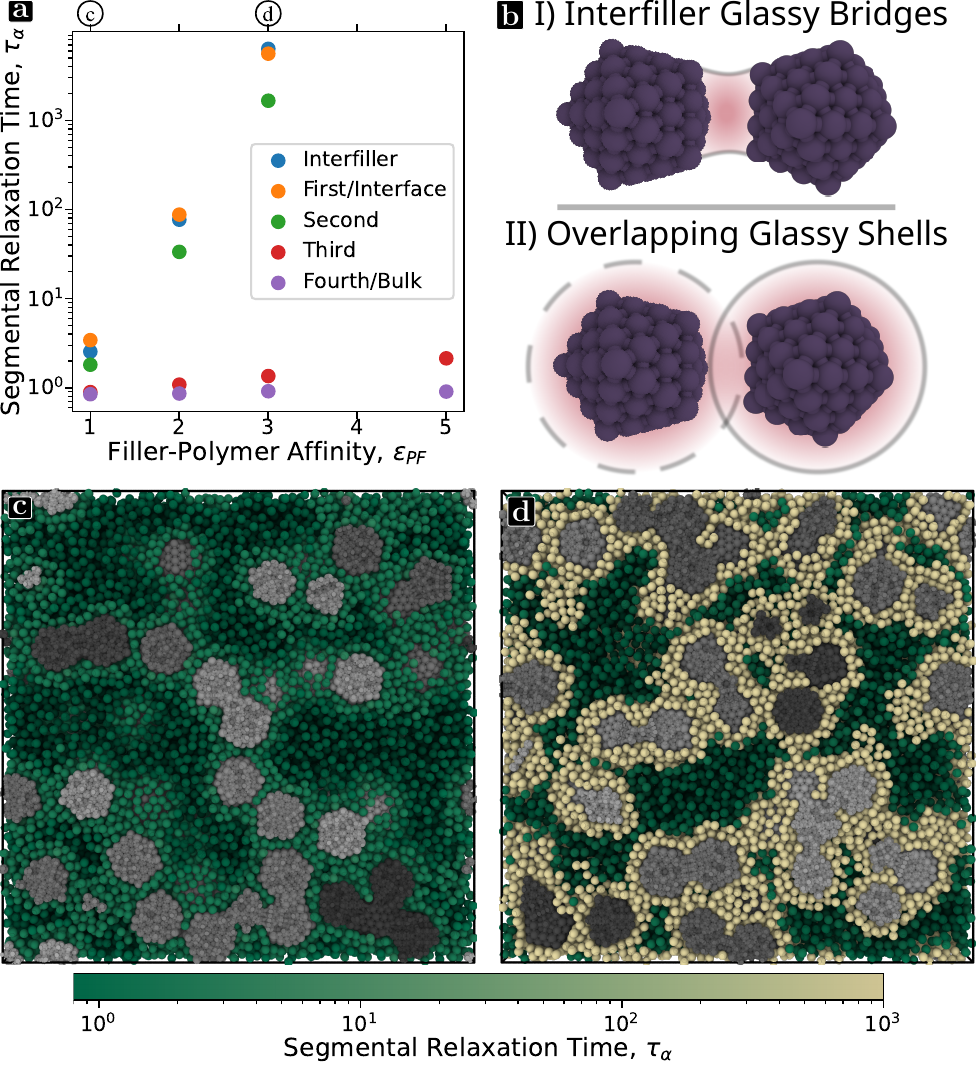}
  \caption{a) Segmental relaxation time versus polymer-filler interaction strength $\epsilon_{PF}$ for the $N_{p}=7$ and 75 PHR filled elastomer system. Scatter points are included for ``Interfiller'' beads, as well as radial shells (1$\sigma$ thick) around the nanoparticle surface. Relaxation time is determined as the time when the self-intermediate structure function decays to a value of 0.2 at a wavenumber of 7.07. b) Schematic illustration of interfiller glassy bridges and overlapping glassy shells. c) Initial configuration for $N_{p}=7$, 75 PHR, and $\epsilon_{PF}=1$. d) Same as c but with $\epsilon_{PF}=3$. Distinct filler clusters are colored with different shades of gray and elastomer beads are colored by relaxation time. e) Color bar mapping relaxation time to color in c and d.}
  \label{fig:logtau-Np7_75phr}
\end{figure}

As indicated by Figure~\ref{fig:logtau-Np7_75phr}a, our systems can be divided into three classes based on the impact of filler particles on polymer segmental relaxation.
When $\epsilon_{PF}=1$, polymer dynamics are essentially unaffected by proximity to filler particles, indicating the absence of glassy shell or bridge effects.
At intermediate interaction strengths ($\epsilon_{PF}=2$ or $3$), polymer dynamics within two segment lengths of the nanoparticle surface (the short range expected far above $T_g$~\cite{Lang2013, Lang2014, Marvin2014, Mangalara2015, Mangalara2016, Riggleman2006, hanakata_local_2012, lang_interfacial_2013, Schweizer2019}) are appreciably slower than the bulk but are not arrested on simulation timescales.
That is, interfacial polymer segments are in an intermediate viscous shell regime.
Finally, at strong interactions ($\epsilon_{PF}=5$), near-particle relaxation times exceed simulation timescales, indicating vitrification and the formation of glassy shells in near-surface polymer.
In all cases, the relaxation time of polymers located between filler clusters (glassy bridges, see Figure~\ref{fig:logtau-Np7_75phr}b) is essentially the same as in near-surface shells.
This indicates that glassy or viscous bridging, to the extent it is present, results from simple overlap of glassy shells rather than from uniquely strong dynamical suppression of interfiller bridging domains.
Other variables (particle loading and structure) exert relatively minor influence on near particle dynamics, such that $\epsilon_{PF}$ is the dominant controlling parameter for shell classification (see Section~\ref{sec:tau} and Figure~\ref{fig:small_strain-logtau} in the Supplementary Information, SI).

The larger-scale implications of these findings can be seen in our initial configurations in Figure~\ref{fig:logtau-Np7_75phr}c and~\ref{fig:logtau-Np7_75phr}d with polymer segments colored by their relaxation time.
At $\epsilon_{PF}=1$, regions near particles exhibit only marginally perturbed dynamics relative to bulk-like regions far from particles (panel c).
In contrast, at high $\epsilon_{PF}$, particles along with their glassy shells form a percolating network throughout the system (panel d), resembling bicontinuous phase-separated domains of bulk elastomer and filler networks connected via immobilized polymer.
While direct particle-particle contacts are not immediately apparent in these 2D projections, particle percolation is expected to be present at our filler loading and structure.
Indeed, prior simulation and experimental studies suggest that the percolation threshold for carbon black-like aggregates is well below our lowest filler volume fraction (26.1\%).~\cite{meier_analysis_2007, coupette_percolation_2021}
Although our preparation protocol maximizes ideal dispersion and minimizes initial particle-particle contacts, mechanical percolation is likely present and may intensify under deformation.
Importantly, particle percolation via direct contacts is known to significantly influence mechanical properties even in the absence of frozen polymer shells, and we expect this influence to be further amplified by the presence of immobilized polymer, which enhances network connectivity and stiffness.

\subsection{Emergence of glassy strain softening and yield at low strain in the presence of strong attractions} \label{sec:results:stress}

\begin{figure}[!htp]
  \centering
  \includegraphics[]{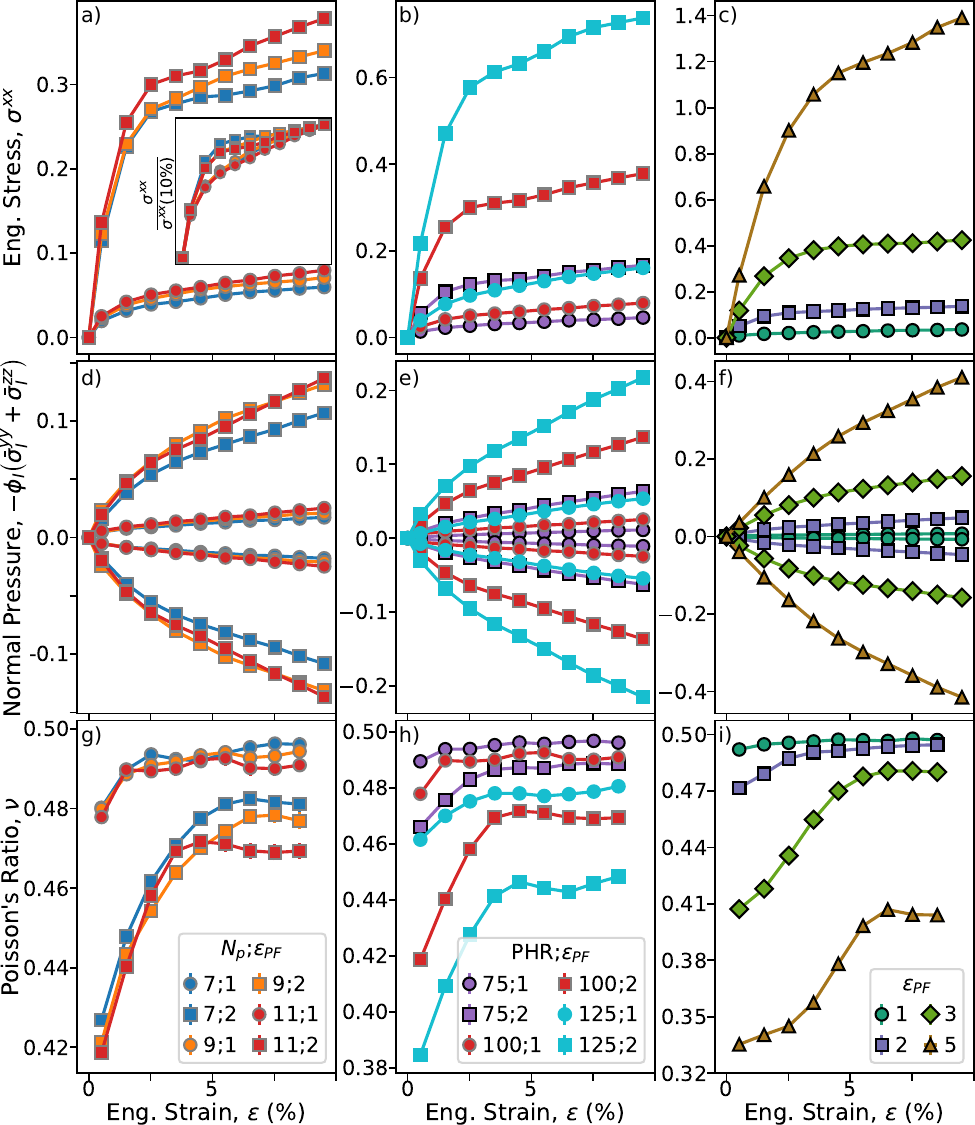}
  \caption{Deformation of filled rubber at different polymer-filler interaction strengths $\epsilon_{PF}$ at (a, d, g) 100 PHR filler loading with various number of primary filler particles in cluster ($N_{p}$), (b, e, h) $N_{p}=11$ with various filler loadings, and (c, f, i) 75 PHR and $N_{p}=7$ with standard error bars from 100 replicates each.
  (a, b, c) Engineering stress $\sigma^{xx}$, (d, e, f) species normal pressure multiplied by species volume fraction (filler is +ve and polymer is -ve), and (g, h, i) Poisson's ratio $\nu$ versus engineering strain $\varepsilon$.}
  \label{fig:linear}
\end{figure}

As expected, increasing polymer-particle attraction enhances mechanical response (Figure~\ref{fig:linear}a-c). More importantly, it drives the emergence of a new feature in the low-strain response regime -- a sharp yield at approximately 2\% strain. While superficially this might be assigned to the Payne effect -- a yield event in filled elastomers that is attributed to the breakdown of a physical filler network under deformation~\cite{Payne1962, Robertson2021} -- in detail this does not appear to be the correct physical assignment. The Payne effect is a broad softening of the modulus that spans a few decades in strain amplitude and is commonly observed most clearly in the 5\% to 10\% strain range.~\cite{Payne1962, Robertson2021} In comparison, the yield observed in our strongly interacting systems is quite sharp and is uniformly at about 2\%, suggesting a distinct yield mechanism.

Rather than emanating from particle network behavior, this yield event, and the ultra-high-modulus-regime that precedes it, must be understood as an unavoidable consequence of the presence of glassy bridges. This can be understood from a formal analysis of the connection between high-frequency glassy relaxation dynamics and stress-strain response via the generalized linear viscoelastic relations.
The elongational stress $\sigma$ is related to the relaxation modulus $E$ via the equation,~\cite{ferry_viscoelastic_1980}
\begin{equation} \label{eq:relaxationmodulus-startup}
  \begin{aligned}
  \sigma \left( t \right)=\dot{\varepsilon }\int\limits_{0}^{t}{E\left( t-s \right)ds}
  \end{aligned}
\end{equation}
where $t$ is the time since deformation began and $\dot{\varepsilon}$ is the (constant) elongational strain rate.

\begin{figure}[!htp]
  \centering
  \includegraphics[]{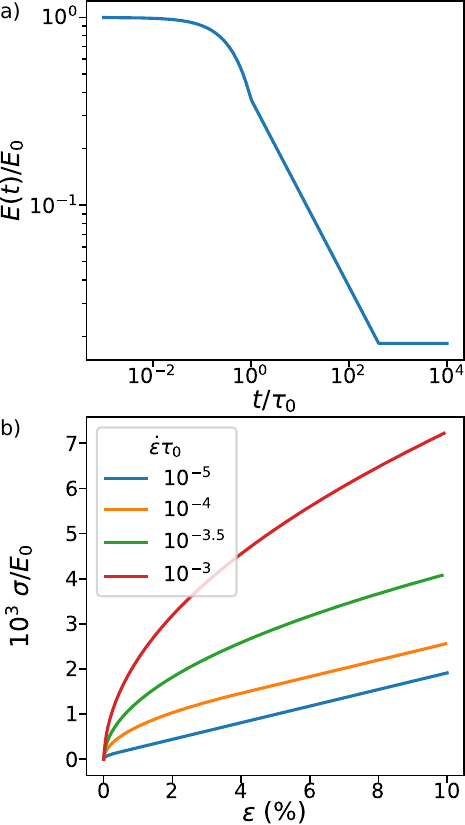}
  \caption{a) Relaxation modulus model (Equation~\ref{eq:elastomerEt} with 20 Kuhn segments, i.e., effective monomers) employed in computing anticipated linear response startup elongation modulus of an elastomer.
          b) Predicted stress versus strain for linear response in startup elongation, using the relaxation modulus in panel a, accounting for short-time integration of high frequency modes via Equation~\ref{eq:relaxationmodulus-startup}.}
  \label{fig:stresstartup}
\end{figure}

For a glass-forming crosslinked polymer, $E(t)$ exhibits an initial glassy plateau, followed by an exponential decay, the Rouse regime, and then the rubbery plateau (see Figure~\ref{fig:stresstartup}a and extended discussion in Section~\ref{sec:derivation} in the SI).
We characterize the rheological response of the material by a Deborah number, defined as the product of its $\alpha$ relaxation time $\tau_0$ and strain rate $\dot{\varepsilon}$ (where $\dot{\varepsilon}\tau_0 \ll 1$, $\dot{\varepsilon}\tau_0 \gg 1$, and $\dot{\varepsilon}\tau_0 \sim 1$ indicate fluid-like response, solid-like response, and intermediate response with comparable relaxation and deformation rates, respectively). 
Numerical solution of Equation~\ref{eq:relaxationmodulus-startup} with this form of $E(t)$ predicts a qualitative change in \emph{linear regime} stress response as $\dot{\varepsilon}\tau_0$ even loosely approaches unity. As shown by Figure~\ref{fig:stresstartup}b, when $\dot{\varepsilon}\tau_0$ approaches $10^{-3}$, a low-strain glassy modulus emerges in the stress response, followed by a softening by around 2\% strain. This is not a genuine yield, but the result of integrating in the glassy and Rouse modes at low strain. This turnover will be further sharpened by actual nonlinear yield of glassy modes at similar strains.~\cite{bowden_yield_1973, papakonstantopoulos_molecular_2008, razavi_crazing_2020}

With this in mind, we proceed by dividing the low-strain response into two regimes and considering them separately:
(1) the \emph{rubbery modulus} at strains beyond 2\%,
and (2) the ultra-low-strain transient glassy response that precedes it.
At strains typically relevant to elastomers (ca. 5\% or greater), stress response in the rubbery regime is most relevant;
the impact of the ultra-low-strain glassy response regime is to generate a vertical offset in the stress response at higher (elastomer-relevant) strains.


\subsection{Glassy bridge effects on the rubbery modulus} \label{sec:results:nu_mismatch}


As shown in Figure~\ref{fig:linear}, the rubbery response of systems with viscous and glassy shells exhibit the same central features of the Poisson's ratio mismatch mechanism as systems without glassy shells.~\cite{Smith2019,Kawak2024} Enhanced stress is mirrored by an enhanced internal normal stress balance between filler and polymer (Figures~\ref{fig:linear}d-f) and reduced Poisson's ratio (Figures~\ref{fig:linear}g-i) -- core signatures of the polymer/filler volume competition that underpins Poisson's ratio mismatch effects.\cite{Smith2019, Kawak2024}

To quantitatively assess the predominance of the Poisson's ratio mismatch mechanism (M4) over other glassy-shell-based reinforcement mechanisms, we compare its predictions for the rubbery moduli to measurements at 5.5\% strain.
The core prediction of M4 relates the composite modulus $E_{c}$ to the change in Poisson's ratio in the composite relative to the neat elastomer, $\nu_{c} - \nu_{e}$:~\cite{Smith2019}
\begin{equation} \label{eq:nu_mismatch}
        E_{c} = f \left[ E_{e} + 2 K_{e} \left( \nu_{e}-\nu_{c} \right) \right]\ ,
\end{equation}
where $f$ is a strain amplification factor (measured based on chain deformation rather than employed as a fit parameter, as described in SI Section~\ref{sec:localization}) and $E_{e}$, $K_{e}$, and $\nu_{e}$ are the Young's modulus, bulk modulus, and Poisson's ratio of the neat elastomer measured at the same strain, respectively.
This equation predicts that adding filler enhances the Young's modulus by invoking a contribution from the elastomer's bulk modulus, with a magnitude controlled by the suppression in composite Poisson's ratio.

\begin{figure}[htp!]
  \centering
  \includegraphics[]{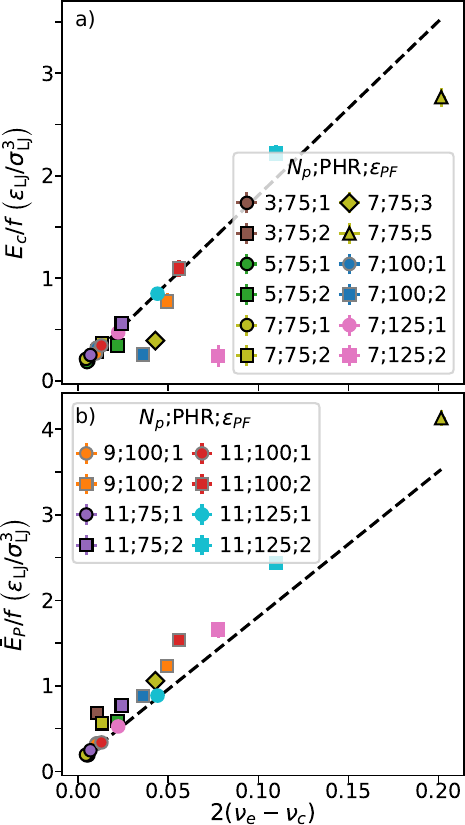}
  \caption{a) Normalized Young's modulus $E_{c}/f$ and b) normalized polymer Young's modulus $\bar{E}_{P}/f$ versus Poisson's ratio mismatch $2\left(\nu_e - \nu_{c} \right)$, all computed at a strain of 5.5\%.
  These moduli are normalized by the strain amplification factor $f$, which is measured from simulation as described in the SI and is not a fit parameter.
  All scatter point error bars are standard errors from 100 replicates.
  The dashed line denotes the slope predicted by the measured bulk modulus of the neat elastomer.}
  \label{fig:theory_sameplot}
\end{figure}

In prior work, we reported that Equation~\ref{eq:nu_mismatch} predicts reinforcement across a wide range of nanoparticle loadings and structures\cite{Smith2019}, in the absence of glassy bridges.
In contrast, as can be seen in Figure~\ref{fig:theory_sameplot}a, for many of the systems with $\epsilon_{PF} > 1$, Equation~\ref{eq:nu_mismatch} \emph{overpredicts} the degree of reinforcement when it is applied to the overall composite stress. However, as shown in Figure~\ref{fig:theory_sameplot}b, it provides an excellent description of the elongational stress within the polymer component of the composite, a la
\begin{equation} \label{eq:nu_mismatch**}
  \bar{E}_{P} = f \left[ E_{e} + 2 K_{e} \left( \nu_{e}-\nu_{c} \right) \right]\ ,
\end{equation}
Here the overbar denotes a partial volumetric component stress within the composite, defined such that the total stress is given by a volume-fraction-weighted sum over partial volumetric component stresses ($\sigma = \phi_P \bar{\sigma}_P + \phi_F \bar{\sigma}_F$, $E = \phi_P \bar{E}_P + \phi_F \bar{E}_F$, see Methods Section for formal definition). On the basis of the physical origin of these equations, this is unsurprising: this relationship is the consequence of compelling the elastomer matrix to deform at a non-native Poisson ratio, without considering \textit{direct} contributions to elongational stress from the filler network. These contributions were negligible in the absence of glassy bridges and could be neglected to leading order; evidently they are significant here.

We can account for direct elongational contributions from the filler network by decomposing the composite modulus into direct contributions from polymer and particle components:
\begin{equation} \label{eq:mod-decomp}
  \begin{aligned}
          E_{c}-E_{e}  = \phi_{P} {\left(\bar{E}_{P}-E_e\right)}  +  \phi_{F} {\left(\bar{E}_{F}-E_e\right)} .
  \end{aligned}
\end{equation}
Substituting Equation~\ref{eq:nu_mismatch**} into Equation~\ref{eq:mod-decomp} then yields a generalized prediction for the modulus of a composite, accounting for multiple distinct mechanisms of reinforcement:
\begin{equation} \label{eq:reinforcement_contributions}
\left( E_c - E_e \right)_{\text{predicted}} =
\underbrace{\phi_P E_e (f - 1)}_{\text{M1: Strain Localization}} +
\underbrace{\phi_F \left( \bar{E}_F - E_e \right)}_{\text{M2: Elongational Glassy Briding}} +
\underbrace{2 \phi_P f K_e (\nu_e - \nu_c)}_{\text{M4: $\nu$ Mismatch}}.
\end{equation}
Here, term 2 corresponds to mechanism 2 -- direct elongational stress contributions from the filler network that were missing in our original model. The first and third terms represent a decomposition of the direct polymer contribution into a term (term 1) quantifying strain localization effects from particles and nearby frozen elastomer (mechanism 1) and a term (term 3) quantifying the Poisson's-ratio-mismatch mechanism (mechanism 4). 

\begin{figure}[!htp]
  \centering
  \includegraphics[]{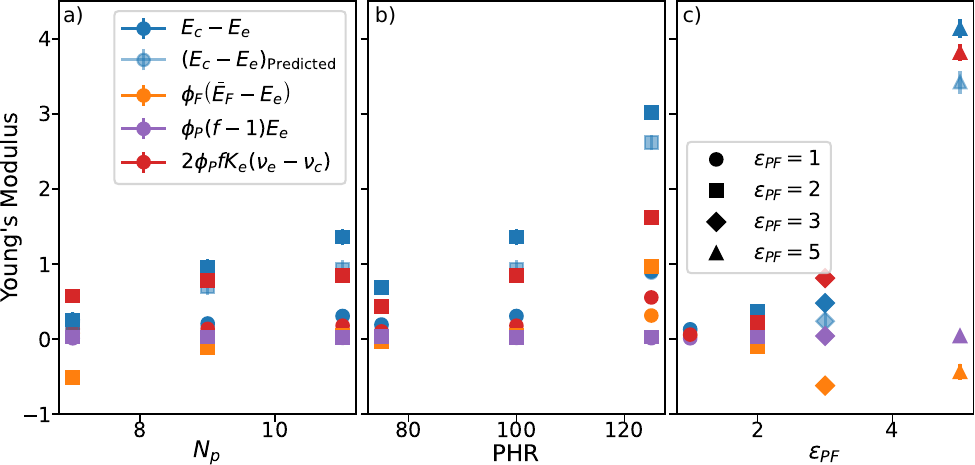}
  \caption{Plots of measured (MD simulation) and predicted (Equation~\ref{eq:reinforcement_contributions}) difference in composite and neat Young's moduli, $E_{c}-E_{e}$, along with the contributing terms in the right-hand-side of Equation~\ref{eq:reinforcement_contributions}.
  All scatter point error bars are standard errors from 100 replicates.
  From left to right, the columns show data for the 100 PHR, $N_{p}=11$, and $N_{p}=7$ + 75 PHR systems, respectively.}
  \label{fig:species_decomp}
\end{figure}

To test Equation~\ref{eq:reinforcement_contributions}, we measure the direct filler elongational contribution (term 2) from simulation, we measure $f$ and $\nu$ from simulations, and terms 1 and 3 then yield predicted values for the strain localization and $\nu$ mismatch mechanisms. As shown in Figure~\ref{fig:species_decomp}, this equation provides a good leading order prediction of composite moduli in our simulated systems. 

Remarkably, as shown in Figure~\ref{fig:species_decomp}, the direct filler contribution to composite stress, $\phi_{F} \left(\bar{E}_{F}-E_e\right)$, is in many cases \emph{negative}. This seems surprising: hypothesized glassy bridging is predicated on the idea that glassy interparticle bridges lend \emph{elongational} cohesion to the nanoparticulate network. However, the reason for this contrary outcome is simple. As discussed above, beyond 2\% strain, the elongational glassy modes within the system have yielded or fully integrated in. \emph{Elongational glassy cohesive effects are thus necessarily lost beyond the ultra-low-strain regime and do not contribute to the rubbery elongational modulus.} This essentially rules out the possibility of mechanism 2 - direct elongational cohesion via glassy bridges - in the rubbery response regime.

Other mechanistic contributions to the stress response are evidently of second or higher order. Strain localization effects, as shown in Figure~\ref{fig:species_decomp}, are of smaller magnitude than these effects and are not leading order. Moreover, where there are second-order deviations of the data from Equation~\ref{eq:reinforcement_contributions}, they generally take the form of a modest underprediction. Returning to the list of postulated potential mechanisms in the introduction, we tentatively assign this second-order underprediction to mechanism 3 - enhanced effective crosslink density due to tethering at the particle interface.~\cite{Sternstein2002, Steck2023, Dhara2022, Dhara2024} This would enter the model through an enhancement of $E_e$ relative to the neat elastomer value.
In any case, the correction is second order.

These results indicate that the Poisson's ratio mismatch between nanoparticle network and elastomer matrix is the dominant mechanism of rubbery modulus reinforcement, including in the presence of glassy bridges. This implies that glassy shell and bridge effects primarily enhance the rubbery modulus beyond 2\% strain by buttressing the particle network in the compressive direction, strengthening the Poisson's ratio mismatch effect.

\subsection{Ultra-low-strain glassy response regime}


Finally, we consider the impact of glassy bridging in the ultra-low-strain regime prior to the $<$2\% glassy softening. By extrapolating the line corresponding to the 5.5\% tangent modulus back to a strain of zero, we extract the vertical offset $C_0$ in the rubbery response (at 5.5\% strain) produced by the ultra-low-strain glassy response.
As shown in Figure~\ref{fig:intercept_vs_slope-small_strain}, $C_0$ is proportional to the tangent modulus of the system at \emph{1.5\% strain}, which confirms that this offset reflects the degree of glassy stiffening in this ultra-low-strain regime.

\begin{figure}[!htb]
  \centering
  \includegraphics[]{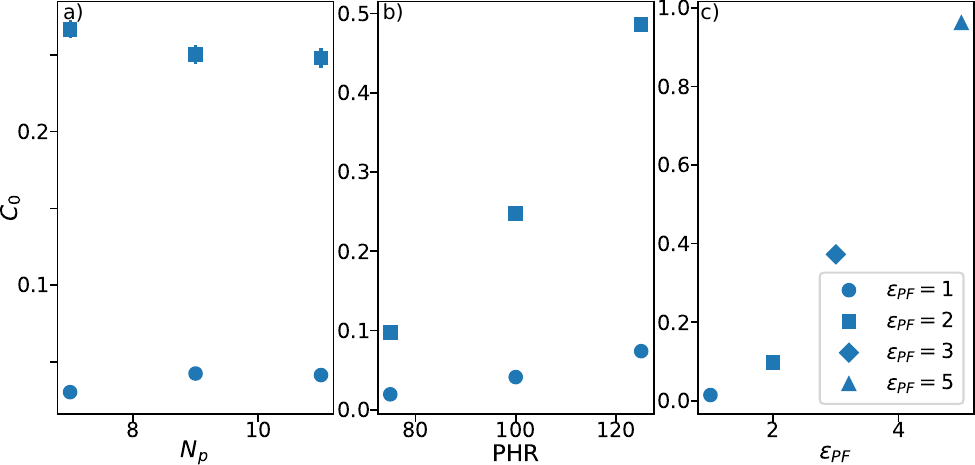}
  \caption{Extrapolation to the intercept of the stress-strain curve from the rubbery modulus at 5.5\% strain ($C_{0} = \left[ \sigma - E_{c} \varepsilon \right]_{\varepsilon=5.5\%}$) versus a) $N_{p}$ at 100 PHR, b) PHR loading at $N_{p}=11$, and c) $\epsilon_{PF}$ at $N_{p}=7$ and 75 PHR.
  All scatter point error bars are standard errors from 100 replicates.
  }
  \label{fig:intercept_vs_slope}
\end{figure}

The behavior of the ultra-low-strain regime, analyzed in this manner, resembles a percolating glassy network that is mechanically diluted by soft surrounding rubbery regions. As shown by Figure~\ref{fig:intercept_vs_slope}, $C_0$ increases strongly with $\varepsilon_{PF}$; this is consistent with the expected emergence of a glassy network as glassy shells emerge. It additionally increases with particle loading; this is consistent with a gel-like dilution model, akin to dilution of crosslinked elastomers by a solvent. In this case, however, the diluted network is glassy rather than rubbery. In contrast, $C_0$ is relatively insensitive to particle structure (Figure~\ref{fig:intercept_vs_slope}a); this is reasonable for a percolated network where structure does not change the fraction of glassy material. 



\section{Discussion and Conclusions}


These findings point to a clear-cut diagnostic for the presence of glassy bridge effects.
Glassy bridges \emph{unavoidably} introduce a glassy response at less than 2\% strain, which then yields to the rubbery response.
This is a physical consequence of introducing a percolating glassy network, diluted by rubbery content, and is formally required by low-strain integration-in of this network's glassy modes. A presence of this regime suggests the presence of glassy bridges; its absence unambiguously denotes their absence.

These results, taken together with our prior findings, provide a unified understanding of the century-old question of the mechanism of low-strain reinforcement in filled elastomers.
While earlier models emphasized mechanisms such as bound rubber, glassy bridging, and transient crosslinking, as \emph{causal} mechanisms of reinforcement, our simulations demonstrate that these effects play quantitatively \emph{augmenting} but not mechanistically \emph{causal} roles.
The causal mechanism of rubbery regime reinforcement is instead a competition between elastomer and particle networks over controlling the system volume under deformation, as reflected in their differing native Poisson's ratios. This competition induces a bulk modulus-mediated resistance to deformation. 
This mismatch is significantly \emph{amplified}, but not \emph{caused} by glassy shells (when present). These shells buttress the nanoparticle network against compressive strain, biasing the composite towards more volumetric expansion and greater contributions from the bulk modulus.

\begin{figure}[!htb]
  \centering
  \includegraphics[]{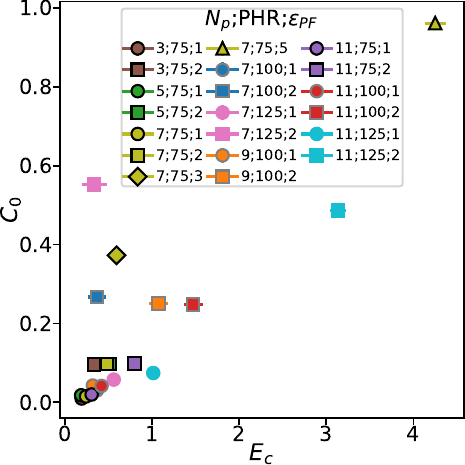}
  \caption{Intercept extrapolated from the rubbery modulus at 5.5\% strain versus rubbery modulus at 5.5\% strain.
  All scatter point error bars are standard errors from 100 replicates.
  }
  \label{fig:intercept_vs_slope-all}
\end{figure}

Finally, this study provides a framework for rational design of low-strain reinforcement in elastomeric nanocomposites via control of particle structure, loading, and interactions: the combination of these variables allows for a degree of independent control of the rubbery modulus versus the ultra-low-strain glassy response.
To illustrate this, in Figure~\ref{fig:intercept_vs_slope-all}, we plot the value of $C_0$ (which characterizes the strength of the low-strain glassy response) versus the rubbery modulus of the composite at 5.5\% strain for all studied systems. As can be seen here, introducing glassy shell and bridge effects (increasing $\epsilon_{PF}$) tends to raise both the rubbery modulus and ultra-low-strain glassy modulus, as does increasing loading (especially when glassy bridges are present).
In contrast, increasing particle structure tends to increase the rubbery modulus without strengthening the ultra-low-strain glassy response.
The combination of these variables thus offers the potential to control the shape of the stress response below 10\% strain in a fairly facile way: filler cluster size and morphology can enhance the rubbery modulus without inducing stress softening related to glassy polymer layers;
increasing filler loading and polymer-filler affinity enhances stiffness but introduces pronounced stress softening at low strains.
This informed control of filler characteristics and polymer-filler interactions provides targeted pathways for designing elastomeric materials tailored to specific performance criteria in demanding industrial environments, such as automotive tires, vibration damping, and impact-resistant applications.


\noindent \textbf{Conflict of Interest}
The authors declare no conflict of interest.

\noindent \textbf{Data Availability Statement}
The data that support the findings of this study are available upon reasonable request.

\begin{suppinfo}


The Supporting Information is available free of charge online.


\end{suppinfo}

\begin{acknowledgement}


This material is based upon work supported by the U.S.~Department of Energy, Office of Science, Office of Basic Energy Sciences, under Award Number DE-SC0022329.


\end{acknowledgement}

\clearpage

\begin{center}
\noindent \textbf{\Huge Supplementary Information: Glassy interphases reinforce elastomeric nanocomposites by enhancing percolation-driven volume expansion under strain}
\end{center}

\setcounter{equation}{0}
\setcounter{figure}{0}
\setcounter{table}{0}
\setcounter{page}{1}
\makeatletter
\renewcommand{\thepage}{S\arabic{page}}
\renewcommand{\thesection}{S\arabic{section}}
\renewcommand{\thetable}{S\arabic{table}}
\renewcommand{\thefigure}{S\arabic{figure}}
\renewcommand{\theequation}{S\arabic{equation}}

\section{Additional results on local relaxation times} \label{sec:tau}

\begin{figure}[htb]
  \centering
  \includegraphics[]{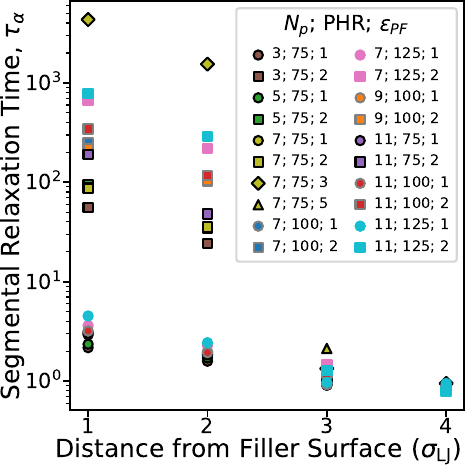}
  \caption{Segmental relaxation time $\tau_{\alpha}$ versus distance (in LJ units) from filler nanoparticle surface for all systems probed. Beads are arranged into bins sized $\sigma_{LJ}$ so that the result at $1\sigma_{LJ}$ includes all polymer beads below a distance of $1\sigma_{LJ}$ away from the nanoparticle center. Legend indicates the filler structure $N_{p}$, loading (in PHR), and polymer-filler interaction strength $\epsilon_{PF}$ for each scatter point. Additionally, systems with different $\epsilon_{PF}$ are distinguished via scatter point shape with $\epsilon_{PF}$ values of 1, 2, 3, and 5 corresponding to cicle, square, diamond, and triangle, respectively. Relaxation time is calculated as the time when the value of the self-intermediate structure function at a wavenumber of 7.07 decays to a value of 0.2. All points have error bars from standard errors of the mean of 100 replicates.}
  \label{fig:small_strain-logtau}
\end{figure}

\section{Quantification of strain localization} \label{sec:localization}

The Poisson's ratio mismatch theory (Eq.~\ref{eq:nu_mismatch}) predicts the composite modulus with a multiplicative scalar correction for strain localization.
This scalar is called the strain amplification factor, $f$.
$f$ is generally $ \ge 1$ and a unit value indicates no amplification.
We quantify this using the end-to-end distance of polymer strands in the stretch direction $R_{ee, x}$ for the neat and composite using the following relationship for a molecular-level polymer strain $\gamma_{x}$ at any strain $\varepsilon$:
\begin{equation}
        \gamma_{x}(\varepsilon) = \ln \left( \frac{\langle R_{ee, x}(\varepsilon) \rangle }{\langle R_{ee, x} \rangle_{\varepsilon=0} }  \right)
\end{equation}
where $\langle \rangle$ indicates an ensemble average of all polymer strands and the denominator is calculated before any strain.
Then, $f$ is computed for any filled system using the ratio of its molecular strain to that of the neat elastomer at the same strain:
\begin{equation}
        f(\varepsilon) = \frac{\gamma_{x}(\varepsilon)}{\gamma_{x, e}(\varepsilon)}
\end{equation}
where the subscript $e$ is used for the neat elastomer.

\begin{figure}[!htb]
  \centering
  \includegraphics[]{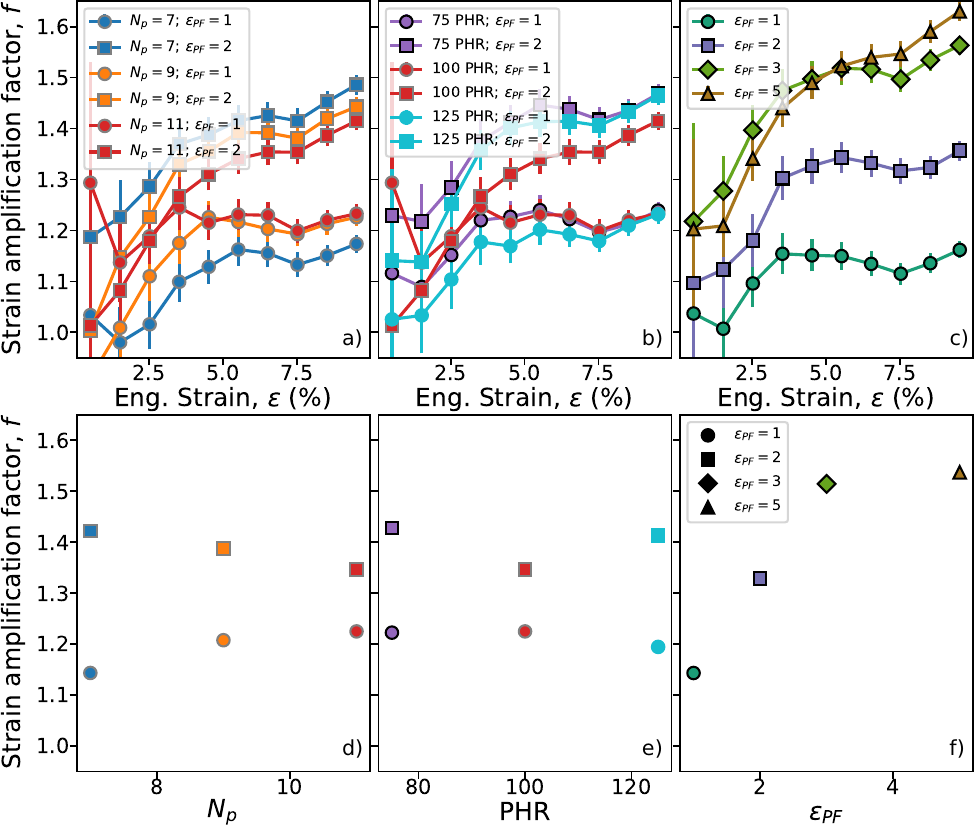}
  \caption{Strain amplification factor data for the different parameter sweeps used in this study. Standard errors of the mean of 50 replicates are used for the error bars.}
  \label{fig:strain_amp}
\end{figure}

Figures~\ref{fig:strain_amp} a, b, and c show how $f$ evolves with strain.
Although the data is somewhat noisy, $f$ plateaus after a few \% strains to a near-constant value.
Accordingly, in all of the calculations in the main text, we use an average of the value of $f$ for all strains $> 3.5\%$.
Figures~\ref{fig:strain_amp} d, e, and f show this average for the different systems employed in this study.
The data shows that the polymer-filler interaction strength has the largest effect on the strain amplification factor.

\section{Detailed derivation of low-strain behavior} \label{sec:derivation}

To understand this sharp, low-strain softening in the presence of glassy shells, we consider the connection between linear glassy relaxation dynamics and the stress-strain response of a material under constant-rate elongation.
In the linear regime, the time-dependent stress $\sigma(t)$ is governed by the material's elongational relaxation modulus $E(t)$ and the applied strain rate $\dot{\varepsilon }\left( s \right)$ as
\begin{equation} \label{eq:relaxationmodulus}
  \begin{aligned}
  \sigma \left( t \right)=\int\limits_{-\infty }^{t}{E\left( t-s \right)\dot{\varepsilon }\left( s \right)ds}\ .
  \end{aligned}
\end{equation}
For startup elongation, this strain rate is constant at positive times and zero at negative times, yielding the equation

To predict the linear-regime elongation response of the elastomer, it is then necessary to inject its time-dependent relaxation modulus.
Qualitatively, the relaxation modulus of a crosslinked elastomer exhibits four regimes:
a short-time glassy plateau; a glassy segmental $\alpha$ relaxation process (exponential or stretched exponential in form); a Rouse regime (power law with slope -1/2) associated with chain motion; and a long-time rubbery plateau.
To model this behavior, we adopt a minimalist piecewise function for $E(t)$ that captures these regimes.
\begin{equation} \label{eq:elastomerEt}
  \begin{aligned}
  E\left( t-s \right)=\left\{ 
  \begin{matrix}
   {{E}_{0}}{{\operatorname{e}}^{-\frac{t-s}{{{\tau }_{0}}}}} & t-s<{{\tau }_{0}}  \\
   E_{0} {\operatorname{e}}^{-1}{{\left( \frac{t-s}{{{\tau }_{0}}} \right)}^{-{1}/{2}\;}} & \tau_{0} < t-s < \tau_{R}  \\
   {{E}_{0}}{\operatorname{e}}^{-1}{{\left( \frac{{{\tau }_{R}}}{{{\tau }_{0}}} \right)}^{-{1}/{2}\;}} & t-s>{{\tau }_{R}} \  \\
  \end{matrix} \right.
  \end{aligned}
\end{equation}
Here $\tau_0$ is the Rouse segmental timescale, which for this qualitative treatment we take to be equal to the segmental $\alpha$ relaxation time $\tau_{\alpha}$, $E_0$ is the glassy modulus, and $\tau_R \sim \tau_0 n^2$ is the Rouse time for a strand of $n$ Kuhn segments between crosslinks.

Evaluating the stress integral in Eq.~\ref{eq:relaxationmodulus} with $E(t)$ from Eq.~\ref{eq:elastomerEt} yields the strain-dependent stress for various dimensionless strain rates $\dot{\varepsilon } \tau_0$.
As shown in Figure~\ref{fig:stresstartup}, when $\dot{\varepsilon } \tau_0$ approaches $10^{-3}$, the stress-strain curve exhibits a pronounced high-modulus regime at low strain, followed by a transition to a softer rubbery response.
Importantly, we emphasize that this prediction is \emph{entirely linear} in nature; i.e., this is not a yield event.
Instead, this low-strain response reflects the high-frequency glassy and Rouse responses of the material `integrated in' to the elongational stress response.
This behavior is expected when the deformation rate approaches the inverse Rouse time of the strand, such that low-strain behavior becomes highly influenced by Rouse and glassy response modes.

In typical experimental elastomers, which are tested at temperatures well above $T_g$ and at low strain rates, the dimensionless rate $\dot{\varepsilon} \tau_0$ is much less than $10^{-4}$.
At such conditions, the high-frequency modes relax quickly, and the stress response reflects only the rubbery plateau.
In our simulations, we use a deformation rate of $5\times10^{-5}/\tau_{\text{LJ}}$ (as low as computationally possible within simulation time constraints) and a segmental relaxation time $\tau_\alpha \sim 1$, yielding $\dot{\varepsilon} \tau_0 < 10^{-4}$.
Thus, as per Figure~\ref{fig:stresstartup}, we expect only a modest contribution from high-frequency modes in the absence of glassy shell or bridge effects.

Why, then, does a much stronger low-strain stiffening and subsequent softening emerge here in systems with glassy shells (as per Figure~\ref{fig:linear})?
In systems with strong polymer–filler interactions, the interfacial polymer becomes dynamically arrested on the timescale of deformation, such that locally $\dot{\varepsilon} \tau_0 \sim 1$.
It follows that \emph{any composite exhibiting glassy bridge/shell effect must produce a pronounced transient glassy response at low strains, diluted by the fraction of mobile (non-glassy) polymer.}

We note that while the analysis above is predicated on linearity, glassy materials also exhibit a sharp nonlinear yield by, at most, 1-2 \% strain.
It follows that, in addition to the predicted linear turnover above, the transition to rubbery response may be further sharpened by the nonlinear yield of glassy domains.
This interpretation is consistent with the abrupt softening observed in our simulations at high $\epsilon_{PF}$ (Figure~\ref{fig:linear}a-c).

\section{Vertical stress offset and modulus at ultra-low strains}

\begin{figure}[!htb]
  \centering
  \includegraphics[]{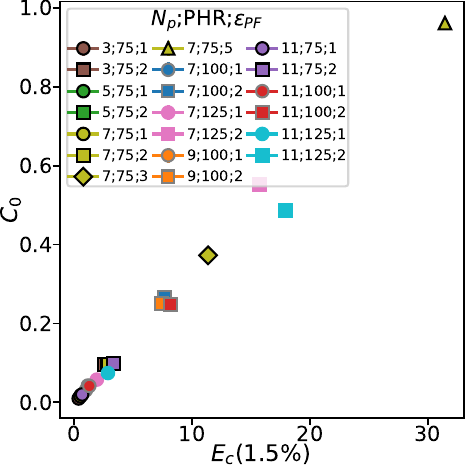}
  \caption{Intercept from rubbery modulus at 5.5\% strain versus the modulus (slope) at 1.5\% strain.}
  \label{fig:intercept_vs_slope-small_strain}
\end{figure}

Figure~\ref{fig:intercept_vs_slope-small_strain} shows the intercept from the rubbery modulus at 5.5\% strain ($C_{0} = \left[ \sigma - E_{c} \varepsilon \right]_{\varepsilon=5.5\%}$) plotted against the slope at ultra-low strain magnitude.
This plot shows a collapse of the points along a straight line, validating the hypothesis that $C_{0}$, the vertical offset of the stress at higher strains, is caused by the initial modulus resulting from glassy contributions to the stress response.

\clearpage
\bibliography{refs}

\end{document}